\documentclass{elsart}

\usepackage{graphics}
\usepackage{graphicx}
\usepackage{epsfig}

\usepackage{amssymb}

\begin{document}

\begin{frontmatter}



\title{Electro-Strong Interaction and High $Q^2$ Events at HERA}


\author{P. C. M. Yock\thanksref{fn1}}
\thanks[fn1]{E-mail: p.yock@auckland.ac.nz}
\address{Faculty of Science, University of Auckland, Auckland, New Zealand}

\begin{abstract}
A previously proposed unified field theory of electro-strong interactions requires two scales of length within hadrons, $\sim10^{-15}m$ and $\sim10^{-18}m$ respectively, and the onset of new phenomena at the shorter scale. Studies at the HERA electron-proton collider at the shorter scale have revealed a possible excess of high-transverse momentum events, as expected, with $Q^2 \geq 30,000$ $(Gev/c)^2$. The collider is currently being upgraded. This will permit a clearer test to be carried out. 
\end{abstract}

\end{frontmatter}

\section{Introduction}
\label{}
The standard model of particle physics is widely regarded as representing the actual structure of matter, because of the diversity of phenomena that are in agreement with it. However, the model includes a large number of empirically determined parameters. It is, therefore, either incomplete, or, as the name suggests, a model only. The inclusion of the phenomenon of confinement in the model renders it especially difficult to confirm or falsify unequivocally, because the confined particles are not able to be studied in isolation.

In this paper an alternative to the standard model is discussed. The alternative was developed about the same time as the standard model (Yock, 1969), and it shares some features with that model. However, it unifies electromagnetic and strong interactions, and is a theory of electro-strong interactions. Its most distinctive requirements are the presence of two scales of length within hadrons, $\sim10^{-15}m$ and $\sim10^{-18}m$ respectively, and the onset of fundamental electro-strong effects at the shorter scale. The electron-proton collider known as HERA at Hamburg affords the first opportunity to search directly for such effects. The present status of the search is discussed here. 

The layout of the paper is as follows. The electro-strong theory is summarized in \S2. A generalized Yuakawa model with two scales of length is presented in \S3. The HERA experiment is discussed in \S4. Conclusions are given in \S5.    

\section{Theory of Electro-Strong Interactions}
The electro-strong theory predates the standard model, yet it shares basic hypotheses with that model. These include the existence of hadron constituents with dual charges, confinement or strong binding by a massless gauge field, charge neutrality of bound states, and the nuclear force being a reidual effect of the confinement interaction (Yock, 1969; Fritzsch \textit{et al.}, 1973; Gross \& Wilzcek, 1973; Weinberg, 1973). In contrast to the standard model, however, the fundamental gauge of the electro-strong theory was taken to be the normal electromagnetic group, $U(1)$. The hadronic constituents were assumed to be heavy and highly charged, and termed 'subnucleons'. 

Four generations of subnucleons, that are not replications of one another, were assumed to exist. Conservation laws of baryon number and strangeness, etc., were shown to follow from electric charge conservation. Weak interactions were included, but only qualitatively. The presence of strong interactions in the theory prevented the author from attempting detailed calculations. The theory is highly conjectural (Yock, 1991).

The theory was formulated in an attempt to construct a field theory that is free of the renormalisation divergences and broken symmetries of conventional theory. It has not been demonstrated by direct calculation that either of these goals has been achieved. Nor is it known if they are achievable (Feynman, 1985). The attempt followed here rests on the assumption that a physical zero exists to the generalized Gell-Mann-Low function for the full theory with weak, strong and electromagnetic interactions included.               

\section{Generalized Yukawa Model} 
Low-energy experiments show no hint of highly electrically charged constituents within the proton. Consequently, a generalised Yukawa model of hadrons was proposed (Yock, 1970). The nucleon, for example, was envisaged as a bare, composite nucleon surrounded by a cloud of composite, virtual pions, as shown schematically in the figure. The size of the bare nucleon was conjectured to be $\sim10^{-18}m$, much smaller than the extent of the pion cloud $\sim10^{-15}m$. With these two scales of length, the presence of the high electric charges would remain hidden at all but the highest energies.

\begin{figure}
\begin{center}
\includegraphics*[width=6cm]{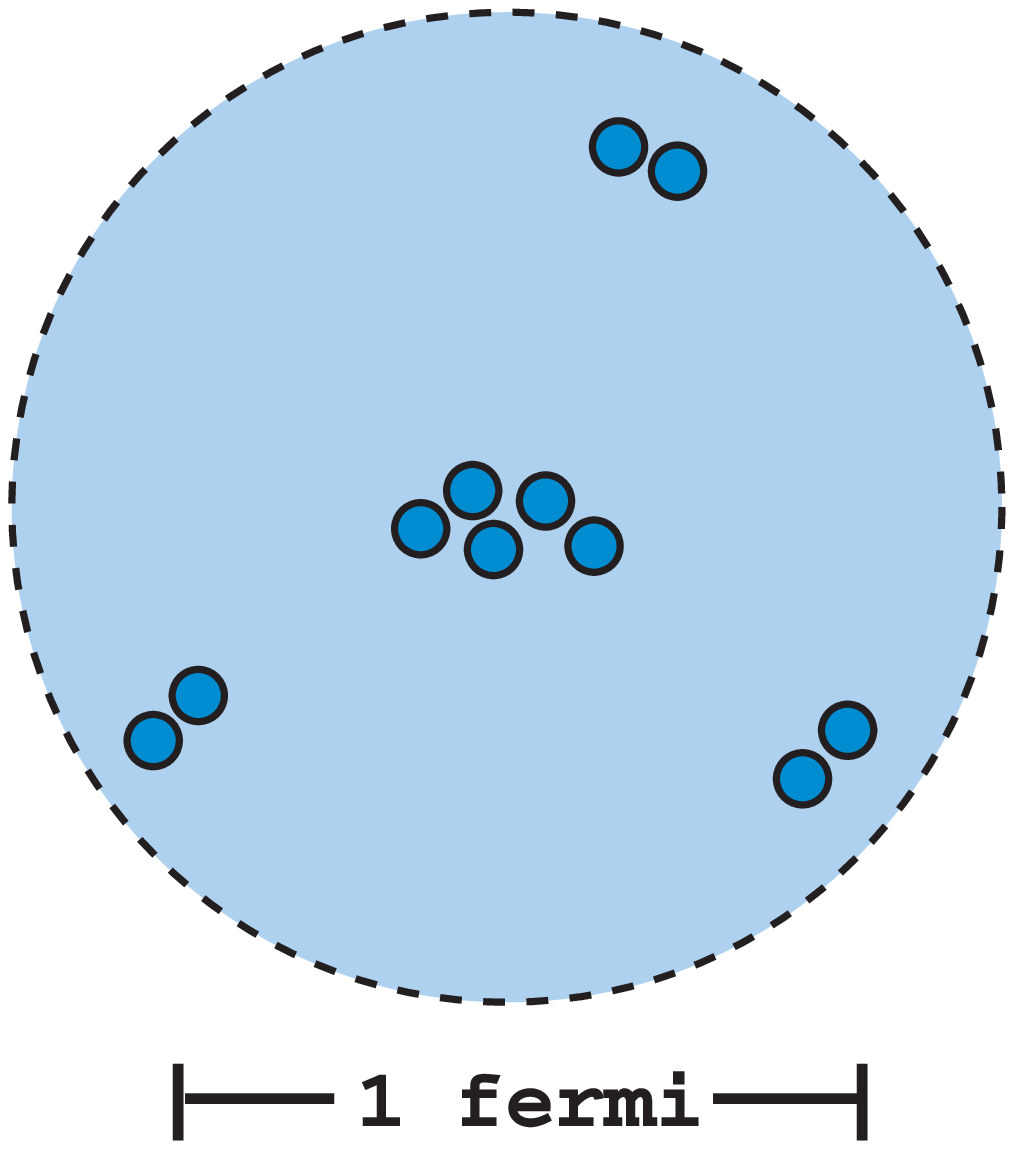}
\end{center}
\caption{Generalized Yuakawa model. The small circles represent tightly bound subnucleons comprising the bare nucleon and virtual pions. The dashed line represents the extent of the pion cloud. In deep inelastic $e-p$ interactions at SLAC energies, the electron is assumed to be scattered by the bare proton as a whole. At HERA energies, deep inelastic interactions with charged mesons in the cloud may occur (see \S3), and also internal structure of the bare nucleon may be resolved (see \S4).}
\label{fig.Yukawa}
\end{figure}

The model requires electrons to be scattered coherently by the bare nucleon at SLAC energies, and that $\nu W_2 \approx (Q^2/2M\nu) \times f(Q^2/2M\nu)$, where $f(x)$ is the probability that the bare nucleon has charge $+e$ and fraction $x$ of the total nucleon mass. This assumes the bare nucleon is heavier than the individual bare mesons, and that it therefore dominates the scattering at SLAC energies (Yock, 1991).

The model is only rudimentary, but it is probably consistent, at least in principle, with the wealth of low-energy nuclear phenomena that have been accounted for in terms of meson exchange models. It may also receive support from high-energy data. Recently, the ZEUS group at HERA reported clear evidence of a high rate of energetic neutron production in the forward direction in deep inelastic $e-p$ interactions (Derrick, \textit{et al.}, 1996; Breitweg, \textit{et al.}, 2001). The fraction of these events is approximately 10\% of all deep inelastic events, and is independent, within statistics, of the energy and momentum transferred to the electron. Also, the energy spectrum of the neutrons shows no variation with these quantities. These observations are directly suggestive of the original Yukawa model. If the electron interacts with a bare meson in the meson cloud whilst leaving the bare nucleon a spectator, then the above results would follow. This was confirmed by the ZEUS group (Derrick \textit{et al.}, 1996).

The above observations are, however, inconsistent with the standard model, at least as it is normally envisaged. This of course holds that the proton consists of valence quarks, sea quarks and gluons confined in a volume of extent $\approx 1 fm^3$. Together with this mix, quark-antiquark pairs comprising virtual mesons are sometimes included (Watson, 1999). Such a structure, in which all the proton's constituents are continuously interacting with one another, and which does not include a spectator component, would seem unlikely to give rise to the observed events. A standard calculation by the ZEUS group (Derrick \textit{et al.}, 1996) confirmed this.

\section{Deep-inelastic electron-proton scattering at HERA energies}
The generalized Yukawa model requires new phenomena to occur in $e-p$ interactions at energies higher than those reached by SLAC. At momentum transfers sufficient to reveal substructure in the bare nucleon, the nature of the interaction should change qualitatively. The scattering would be caused by the individual subnucleons, and the cross-section determined by their high charges. An excess of events should occur in comparison to any parameterisation of the data that does not include substructure, such as the standard model. As the threshold is approached, form factor effects would be expected to occur. 

Possible evidence for such behaviour was reported recently by both the H1 and ZEUS collaborations at HERA (Adloff \textit{et al.}, 1997; Breitweg \textit{et al.}, 1997). These groups reported an excess of events at momentum transfers $Q^2 \geq 30,000$ $(GeV/c)^2$ in $e^{+}-p$ interactions, corresponding to the onset of a new interaction between positrons and constituents of the proton that acts over a range $\leq 10^{-18}m$ with a strength intermediate between electromagnetic and strong interactions. The excess was, however, at a marginal level of confidence, and it was further reduced with continued running (Adloff \textit{et al.}, 2001; Schneider, 2002).

Clearly, further data are required. The luminosity of the HERA collider is presently being increased by an order of magnitude. This should suffice to clearly reveal substructure in the proton if it is present as proposed above. Ideally, the centre-of-mass energy should also be increased. This could be achieved by combining the LHC and LEP machines at CERN to form a giant $e-p$ collider.             

\section{Conclusions}
Theories of the structure of the proton have evolved over the years. The structureless model of Rutherford was followed successively by Yukawa's meson model, and then a series of quark models of ever-increasing complexity (Watson, 1999). The transition from the Yukawa model to the modern quark models represented a clean break, possibly too clean.

Hybrid models that combine features of the modern theories of hadron structure, and the older theory of Yukawa, are conceivable. Such models appear to have been considered at the time of the original SLAC experiments, albeit briefly, (Mehra, 1994). A general feature of such models is likely to be the presence of two scales of length within hadrons. Both high-energy and low-energy data exist that support the Yukawa model.

One specific hybrid model was discussed here. This is based on a scheme of electro-strong interactions, and it requires the presence of two scales of length. It also requires an excess of high transverse momentum events in deep inelastic $e-p$ interactions at centre-of-mass energies of order some hundreds of $GeV$ in comparison to models with only one scale. A possible hint of an excess has already been reported at the HERA $e-p$ collider. The collider is presently being upgraded. This will permit a clearer test to be carried out. 

The electro-strong theory is an alternative to the standard model, and as such it assumes that the quark model is a model only, i.e. that quarks are not real particles. This is not the first time the suggestion has been made (Gell-Mann, 1964; Richter, 1995). Neither are the HERA data the first to be problematical for the standard model. The spin of the proton, and the lifetimes of the strange particles, for example, have long been unexplained (Bass \& Roeck, 2002; Cheng, 1989). Also, the Higgs boson has not been detected. Conceptual problems, such as fractional charge, replication symmetry and symmetry breaking are also present. The electro-strong theory represents one attempt to avoid such concepts.                    

\section{References}
Adloff, C. \textit{et al.} (1997). \textit{Z. Phys. C} 74, 191.\\
Adloff, C. \textit{et al.} (2001). \textit{Eur. Phys. J. C} 19, 269.\\
Bass, S.D. \& De Roeck, A. (2002). \textit{Nucl. Phys. Proc. Suppl.} 105.1.\\
Breitweg, J. \textit{et al.} (1997). \textit{Z. Phys. C} 74, 207.\\
Breitweg, J. \textit{et al.} (2001). \textit{Nucl. Phys. B} 596, 3.\\
Cheng, H. -Y. (1989). \textit{Int. J. Mod. Phys. A} 4, 495.\\
Derrick, M. \textit{et al.} (1996). \textit{Phys. Lett. B} 384, 388.\\
Feynam, R.P. (1985). \textit{QED}, Princeton University Press, Ch. 4.\\
Fritzsch, H., Gell-Mann, M. \& Leutwyler, H. (1973). \textit{Phys. Lett. B} 47, 365.\\
Gell-Mann, M. (1964). \textit{Phys. Lett.} 8, 214.\\
Gross, D., \& Wilczek, F. (1973). \textit{Phys. Rev. D} 8, 3633.\\
Mehra, J. (1994). \textit{The Beat of a Different Drum}, Oxford University Press, Ch. 23.\\
Richter, B. (1995). \textit{Science} 267, 1612.\\
Schneider, M. (2002). \textit{ArXiv:hep-ex/}0201042.\\
Watson, A. (1999). \textit{Science} 283, 472.\\
Weinberg, S. (1973). \textit{Phys. Rev. Lett.} 31, 494.\\
Yock, P.C.M. (1969). \textit{Int. J. Theor. Phys.} 2, 247.\\
Yock, P.C.M. (1970). \textit{Ann. Phys. (N.Y.)} 61, 315.\\
Yock, P.C.M. (1991). \textit{Particle World} 2, 87.\\
\end{document}